# Direct Photocurrent Detection of Optical Vortex Based on the Orbital Photo Galvanic Effect: Progress, Challenge and Perspective

Jinluo Cheng*, Dehong Yang, Weiming Wang, Chang Xu, Zipu Fan, Dong Sun*

Prof. Jinluo Cheng and Prof. Dong Sun

School of Physics and Laboratory of Zhongyuan Light, Zhengzhou University, Zhengzhou, China

Prof. Jinluo Cheng and Weiming Wang

GPL Photonics Laboratory, State Key Laboratory of Luminescence Science and Technology, Changchun Institute of Optics, Fine Mechanics and Physics, Chinese Academy of Sciences, Changchun 130033, China

E-mail: jlcheng@ciomp.ac.cn

Prof. Jinluo Cheng and Weiming Wang

University of Chinese Academy of Science,

Beijing 100049, China

Dehong Yang, Chang Xu, Zipu Fan and Prof. Dong Sun

International Center for Quantum Materials,

School of Physics, Peking University,

Beijing 100871, China

E-mail: sundong@pku.edu.cn

Prof. Dong Sun

Collaborative Innovation Center of Quantum Matter,

Beijing 100871, China

Prof. Dong Sun

Frontiers Science Center for Nano-optoelectronics,

School of Physics,

Beijing 100871, China

**Abstract**

A photodetector that can directly distinguish the orbital angular momentum (OAM) of light is highly desirable for integrated on-chip OAM detection and focal plane array devices. The recent development of OAM detectors based on the intrinsic orbital photo galvanic effects (OPGE) of materials provide a new route for direct OAM detection that is on-chip scalable with high resolution and speed. In this paper, we summarize the current progress in direct photodetection of OAM via OPGE. We begin with a short review of the basic operation scheme of the OAM detector and provide a comprehensive symmetry analysis to sort out the favorable characteristics of the materials, incorporating considerations from device schemes based on various device performance characteristics and specific application circumstances. From that, we review the current experimental progress and technical challenges, then oversee the possible solutions to these challenges and provide a perspective on the future opportunities of this OAM detection route.

**Introduction**

Since the vortex solutions of the Maxwell–Bloch equations were found and the concept of optical vortex (OVs) was introduced[1] in 1989, OVs technology has developed rapidly. Specifically, the orbital angular momentum (OAM) of light was experimentally discovered by Allen et al. in OVs with helical wavefronts[2], thereby opening a new area of research[3-6]. OVs with OAM are widely used in various applications, ranging from optical manipulation[7,8] to machining[9,10], imaging[11,12], optical communications[13,14], quantum entanglement[15,16], and even astronomical surveys[17,18]. The generation[19-23], manipulation, and detection[24-29] of OVs lay the foundation for various OAM-based technologies. Among them, the detection technology of OAM is indispensable for any OAM-related application. Conventional OAM measurement technologies usually rely on counting the stripes and lattices in the special interferogram and diffraction patterns to extract the phase information of OAM[27,30], which requires bulky and complex setups, thereby imposing a fundamental limit on realizing a miniaturized, integrable and fast operating detector of OAM. The recent rapid development of OAM light sources and manipulation at the nanoscale has enabled the on-chip application of OAM, greatly promotes the demand for on-chip photodetection of OAM[31], especially with direct electrical readout of OAM [Fig. 1a]. Equally important is the application circumstance whenever focal plane imaging of a target with OAM is needed [Fig. 2b]. In such applications, a large number of focal plane arrays of OAM detectors with electric readouts that are compatible with electric readout circuits are needed, with additional requirements for operation speed and unit cell size to achieve high-resolution and high-speed operation to acquire clear image of a high-speed target. However, such on-chip electrical photodetection of OAM remains challenging with very limited prototype solutions.

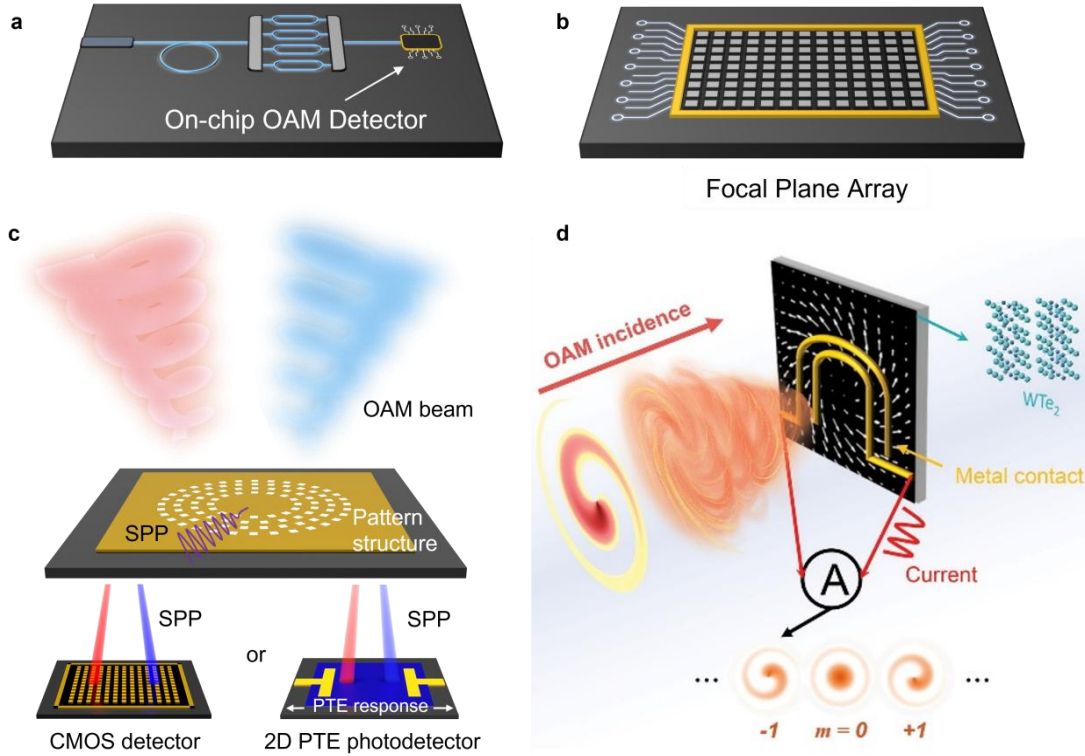

**Fig. 1: Schematic of two typical application perspectives and two typical technical pathways of direct OAM photodetection.** (a-b) Schematic of the direct OAM detector used for on-chip OAM circuits (a) and focal plane arrays (b). (c-d) Schematic of the SPP-based (c) and OPGE-based (d) OAM photodetection. Reproduced with permission[32], Copyright 2020, American Association for the Advancement of Science.

To date, there are only two independent technical paths toward on-chip electrical photodetection of OAM. The first pathway is based on surface plasmon polaritons (SPPs), which relies on well-designed plasmonic nanostructures to separate various OAM modes into focused SPPs at different spatial positions[33-38]. Then, different OAM modes are resolved by detecting the SPPs. The detection usually requires far-field detectors to characterize the transferred SPPs or near-field scanning microscope to record the focused SPPs, which limits system-level integration. Several different schemes have been proposed to solve this issue. For example, by integrating a holographic coupler with a commercial photodiode, certain OAM modes can be detected directly, but others cannot be effectively distinguished[35]. More recent work has used a spin–Hall SPP coupler to sort out different OAM modes and then detect them via the thermoelectric response of $PdSe_2$, which can distinguish different OAM modes simultaneously and demonstrate direct on-chip detection of spin and orbital angular momentum, as well as the chirality and ellipticity of scalar vortex light [Fig. 1c][39]. The second pathway is based on the orbital photogalvanic effect (OPGE) of selected materials, which is the focus of this paper. This orbital photogalvanic effect is driven by the helical phase gradient of light OAM through the electric quadrupole and magnetic dipole response of materials, and then the OAM-dependent photocurrent response component can be extracted from the measurement of the circular polarization-dependent photogalvanic response. The magnitude of the extracted OPGE is proportional to the quantized OAM mode number, which resolves the OAM order of the incident light. Such scheme was first proposed and realized in the near-infrared wavelength range by Ji *et al*. using the type-II Weyl semimetal WTe$_2$ in 2020 [Fig. 1d][32]. After this conceptual demonstration, the recent developments have expanded the functional wavelength range and significantly increased the operation speed[40,41]. Moreover, such detection scheme was recently found to be not limited to Weyl semimetals, a wide range of materials also work similarly with pronounced OPGE responses, even in the challenging mid-infrared wavelength range[42]. A comparison of these detection methods is listed in Table 1.

|  | Traditional methods[27,30] | SPP-based methods | | OPGE-based method[42] |
|---|---|---|---|---|
|  |  | Near field image[35] | Spin-Hall SPP coupler[39] |  |
| Working principle | interferogram and diffraction patterns | well designed metasurfaces to convert OAM modes into focused SPPs at different spatial resolution | detecting the focused SPP using thermoelectric response of PdSe$_2$ | specially designed electrodes to collect the OAM order quantized photocurrent |
| Wavelength | flexible | 633 nm | 8 μm | 4 μm |
| Device footprint | bulky | few hundreds of μm | 200×200 μm | 40×40 μm |
| Operation Speed | very slow | depending on detector | 69/25 μs | 1 ms |
| OAM order | flexible | specific | [-4, 4] | [-4,4] |
| Responsivity | / | tens of $\mu$A/W | 0.2 nA/W | 151.4 nA/W |
| System-level integration | N | Y | Y | Y |

Table 1 A comparison of different OAM detection methods.

With recent rapid developments, such direct OAM detection scheme based on OPGE promises unprecedented opportunities for further development toward scalable, high-resolution, high-speed, on-chip direct OAM photodetectors. In this work, we provide a perspective on the future development opportunities of this technology pathway based on a comprehensive overview of the current technical challenges, and oversee the possible solutions to these challenges and opportunities in the future. We start with a short review of the basic operation scheme of detector based on OPGE and recent experimental progresses. Then the paper features a comprehensive symmetry analysis of all crystal symmetries and then sorts out the favorable characteristics of the material candidates from a symmetry point of view while incorporating the consideration of device scheme based on various device performance characteristics and specific application circumstances, which has not been published elsewhere yet. Based on these results, we discuss in detail how the symmetry of the electrodes affects the characterization of the OAM response current as well as the signal resolution; then, we review the experimental progress on various devices made from different functional materials and the recent progress on improving the operation speed using photoelastic modulator. Furthermore, a new OAM detection strategy by adjusting the linear polarization of the LG beams is proposed. Finally, we discuss the challenge and perspective on the direct detection of OAM order.

**OAM detection via OPGE: operation mechanism, device scheme and experimental progress**
In this section, we first show theoretically how the OAM of light interacts with OAM-sensitive materials and contributes to an OAM-dependent photocurrent response, known as OPGE. Furthermore, we discuss the typical device scheme and operation mechanism based on OPGE.

**Operation mechanism**

Though the structured light can have significant effects on the photocurrent due to the peculiar spatial distribution of the light intensity[43,44], the direct detection of light OAM through the photocurrent response is usually difficult because the photocurrent response does not inherently carry phase information of the incident light, while the OAM of light reflects a peculiar phase variation in the electric field. Except in special nanostructures[43,44] or experimental conditions[45], the light OAM degree of freedom is difficult to interact with electron's "internal" degree of freedom[46,47], e.g., the band index in a crystal, because of the huge scale mismatch between the micrometer scale of light OAM and the nanometer scale of electron's wave function; thus, the electron state can only see the local field of the structured light, instead of its global phase properties. In the approximation of electric dipole interactions, the photocurrent response corresponds to conventional second-order photogalvanic effect, which is determined by the local intensity and polarization of light, irrelevant to the OAM order. The photocurrent response to light OAM can emerge from the phase gradient of optical fields through interactions with the electric quadrupole and magnetic dipole. The spatial variation of OAM light is on the order of wavelength ($\sim \mu m$), which is usually much larger than the scale of the unit cell ($\sim Å$). The several-order length scale differences limit the microscopic interaction between the OAM and solid materials. Theoretically, for a light field $E(r,t) = E(r,\omega)e^{-i\omega t} + c.c.$ interacting with materials, the spatial distribution of the generated DC current can be described as:

$$j_a^{dc}(r) = 2\mathrm{Re}[\alpha_{abc}(\omega)E_b(r,\omega)E_c^*(r,\omega)] + 4\mathrm{Im}\left[\beta_{abcd}(\omega)\frac{\partial E_b(r,\omega)}{\partial r_d}E_c^*(r,\omega)\right] \quad (1)$$

where the subscripts $a, b, c, d$ denote the directions in the Cartesian coordinate system; the response coefficient $\alpha_{abc}(\omega)$ is a third-rank tensor, which describes the conventional second-order response resulting from the dipole approximation; and the response coefficient $\beta_{abcd}(\omega)$ is a fourth-rank tensor, which describes the effects from the gradient of the electric field through the electric quadrupole and magnetic dipole interaction. The response coefficients $\alpha_{abc}(\omega)$ and $\beta_{abcd}(\omega)$ are connected to the general second-order optical conductivities $\sigma_{abc}(q,\omega; q', -\omega)$ in the long wavelength limit $q, q' \sim 0$ as:

$$\sigma_{abc}(q,\omega; q', -\omega) = \alpha_{abc}(\omega) + q_d\beta_{abcd}(\omega) + q'_d\beta_{acbd}(-\omega) \quad (2)$$

The $\beta_{abcd}$ term in Equation (1) describes a part of the spatially nonlocal current that is related to the OAM of light. To elaborate the effect of the helical phase gradient of light OAM in the response, we write the profile of the light field carrying OAM propagating along the $z$ direction in the cylindrical coordinate system:

$$E(r,\omega) = E_0 u_{p,|m|}(\rho, z) e^{im\theta}\frac{\hat{x} + \sigma\hat{y}}{\sqrt{1+|\sigma|^2}} \quad (3)$$

with $\sigma = \sigma_r + i\sigma_i$ describing the light elliptical polarization, $r = \rho\cos\theta\,\hat{x} + \rho\sin\theta\,\hat{y} + z\hat{z}$ and

$$u_{p,|m|}(\rho,z) = \frac{C_{pm}}{w(z)}w_0\left(\frac{\sqrt{2}\rho}{w(z)}\right)^{|m|} L_p^{|m|}\left(\frac{2\rho^2}{w^2(z)}\right)\exp\left[-\frac{\rho^2}{w^2(z)} + i(2p+|m|+1)\eta(z) - i\frac{k\rho^2}{2q(z)}\right] \quad (4)$$

where $E_0$ is the electric field amplitude, $C_{pm} = \sqrt{2p!/\pi(p+|m|)!}$ is a normalization coefficient, $w_0$ is the waist-spot radius of the Gaussian beam, $w(z) = w_0\sqrt{1+z^2/z_0^2}$ is the spot radius at

position $z$, $m$ is the angular quantum number or OAM order number, $p$ is the radial quantum number, $L_p^m(x) = \frac{x^{-m}e^x}{p!}\frac{d^p}{dx^p}(e^{-x}x^{p+m})$ is the generalized Laguerre polynomial, $z_0 = \pi w_0^2/\lambda$ is the Rayleigh length, $\eta(z) = \arctan(z/z_0)$ is the Gouy phase, and $q(z) = z + z_0^2/z$ is the radius of curvature. Owing to the spatial derivative in Equation (1), the OAM order $m$ appears in the response for light field carrying OAM, providing an OAM-dependent photo galvanic response. The spatial distribution of the photocurrent $\boldsymbol{j}^{dc}(\boldsymbol{r})$ generated by the light field in Eq. (3) can be sorted according to the parity dependence of $m$ and $\sigma$ as:

$$j_a^{dc}(\mathbf{r}) = m\left[\sigma_i j_a^{(1)}(\mathbf{r}) + \sigma_r j_a^{(5)}(\mathbf{r})\right] + m j_a^{(2)}(\mathbf{r}) + \left[\sigma_i j_a^{(3)}(\mathbf{r}) + \sigma_r j_a^{(6)}(\mathbf{r})\right] + j_a^{(4)}(\mathbf{r}) \quad (5)$$

where $j_a^{(n)}(\mathbf{r})$ with $n = 1, 2, \cdots, 6$ are given as:

$$\begin{aligned}
j_a^{(1)} &= \operatorname{Im}\left[(\beta_{axyy} - \beta_{ayxy})f_1 g_1 + (\beta_{ayxx} - \beta_{axyx})f_1 g_2\right], \\
j_a^{(2)} &= \operatorname{Re}\left[(\beta_{axxy} + |\sigma|^2 \beta_{ayyy})f_1 g_1 - (\beta_{axxx} + |\sigma|^2 \beta_{ayyx})f_1 g_2\right], \\
j_a^{(3)} &= \operatorname{Im}\left[(\alpha_{axy} - \alpha_{ayx})f_0 g_0\right] + \operatorname{Re}\left[(\beta_{ayxx} - \beta_{axyx})f_2 g_1 - (\beta_{axyy} - \beta_{ayxy})f_2 g_2\right], \\
j_a^{(4)} &= \operatorname{Re}\left[(\alpha_{axy} + \alpha_{ayx})f_0 g_0\right] + \operatorname{Im}\left[(\beta_{axxx} + |\sigma|^2 \beta_{ayyx})f_2 g_1 + (\beta_{axxy} + |\sigma|^2 \beta_{ayyy})f_2 g_2\right], \\
j_a^{(5)} &= \operatorname{Re}\left[(\beta_{axyy} + \beta_{ayxy})f_1 g_1 - (\beta_{axyx} + \beta_{ayxx})f_1 g_2\right], \\
j_a^{(6)} &= \operatorname{Re}\left[(\alpha_{axx} + |\sigma|^2 \alpha_{ayy})f_0 g_0\right] + \operatorname{Im}\left[(\beta_{axyx} + \beta_{ayxx})f_2 g_1 + (\beta_{axyy} + \beta_{ayxy})f_2 g_2\right].
\end{aligned} \quad (6)$$

In Eq. (6), the position dependences of $j_a^{(n)}(\mathbf{r})$, $f_j(\rho, z)$, and $g_j(\theta)$ with $j = 0,1,2$ are not shown explicitly; the functions $f_j$ and $g_j$ depend solely on the radial coordinate $\rho$ and the azimuthal angle $\theta$ separately, and they can be written as:

$$\begin{aligned}
f_0(\rho, z) &= \frac{2|E_0|^2 |u_{p,|m|}(\rho, z)|^2}{1 + |\sigma|^2}, & f_1(\rho, z) &= \frac{2}{\rho}f_0(\rho, z), & f_2(\rho, z) &= \frac{2 f_0(\rho, z)}{u_{p,|m|}(\rho, z)}\frac{\partial u_{p,|m|}(\rho, z)}{\partial \rho}, \\
g_0(\theta) &= 1, & g_1(\theta) &= \cos\theta, & g_2(\theta) &= \sin\theta.
\end{aligned} \quad (7)$$

The current distribution $j_a^{dc}(\mathbf{r})$ in Eq. (5) has the following features:

(1) All terms of $j_a^{(n)}$ depend on $\sigma$ through the factor $1/(1 + |\sigma|^2)$ and show a complicated dependence on m through functions $f_0$, $f_1$, and $f_2$; however, all of them are even functions of $m$ and $\sigma$, which can be observed from Eqns. (6,7).

(2) The photocurrent distribution $j_a^{dc}(\mathbf{r})$ in Eqn. (5) is divided into six terms, each of which is an even or odd function of the OAM order $m$ and the elliptical polarization $\sigma$: the term $m\sigma_i j_a^{(1)}(\boldsymbol{r})$ is an odd function of both $m$ and $\sigma_i$ but an even function of $\sigma_r$; thus, the current direction from this term reverses as one of $m$ and $\sigma_i$ switches signs; the term $m\sigma_r j_a^{(5)}(\boldsymbol{r})$ is an odd function of both $m$ and $\sigma_r$ but an even function of $\sigma_i$; the term $m j_a^{(2)}(\boldsymbol{r})$ is an odd function of $m$ but an even function of both $m$, $\sigma_r$ and the term $\sigma_i$; the term $\sigma_i j_a^{(3)}(\boldsymbol{r})$, is an odd function of $\sigma_i$ but an even function of both $m$ and $\sigma_r$; the term $\sigma_r j_a^{(6)}(\boldsymbol{r})$ is an odd function of $\sigma_r$ but an even function of both m and $\sigma_i$; and the term $j_a^{(4)}(\boldsymbol{r})$ is an even function of $m, \sigma_r, \sigma_i$.

(3) The signs of $m$, $\sigma_r$ and $\sigma_i$ are experimentally controllable, and the contributions of these six terms to $j_a^{dc}(\mathbf{r})$ can be distinguished via a series of experiments by changing the signs of $m$, $\sigma_r$ and $\sigma_i$ only. For example, the CPGE measurement provides the difference in the photocurrent responses between left circularly polarized light ($\sigma_r = 0, \sigma_i = +1$) and right circularly polarized light ($\sigma_r = 0, \sigma_i = -1$), which is contributed by the terms $m j_a^{(1)}(\boldsymbol{r}) + j_a^{(3)}(\boldsymbol{r})$; alternatively, the usual OPGE measurement provides the difference between the CPGE current for OAM order m and OAM order −m, which is contributed by the term $m j_a^{(1)}(\mathbf{r})$.

According to Eqns. (5, 6), the minimum requirement for a certain material to have a possible OPGE response is to have nonvanishing coefficients of $\beta_{axyy} - \beta_{ayxy}$ or $\beta_{ayxx} - \beta_{axyx}$ for $a = x, y$. In fact, such conditions can be easily fulfilled for any crystal because, from the symmetry point of view, the tensor components $\beta_{xxyy}$, $\beta_{xyxy}$, $\beta_{yyxx}$, and $\beta_{yxyx}$ are always nonzero for any point group, and the spatial distribution of $j_a^{dc}(r)$ does not vanish. However, on one hand, the symmetry tells whether there are nonvanishing $\beta_{abcd}$ terms or OPGE terms, it does not tell the response amplitude; on the other hand, there are additional limits imposed from practical considerations, such as device structures and background noise. If the electrodes are considered further, for metallic materials, the current $I(m, \sigma)$ collected by a certain electrode geometry according to the Shockley–Ramo theorem[48-50] can be written as:

$$I(m, \sigma) = \int dr \boldsymbol{j}^{dc}(\boldsymbol{r}) \cdot \boldsymbol{e}(\boldsymbol{r}) = m\sigma_i I^{(1)} + mI^{(2)} + \sigma_i I^{(3)} + I^{(4)} + m\sigma_r I^{(5)} + \sigma_r I^{(6)}$$
$$I^{(n)} = \int dr \boldsymbol{j}^{(n)}(\boldsymbol{r}) \cdot \boldsymbol{e}(\boldsymbol{r}) \tag{8}$$

where $\boldsymbol{e}(\boldsymbol{r})$ is an auxiliary weighting field determined by solving Laplace's equation within a specific device geometry, and in nonmagnetic materials, it is proportional to the electric field distribution when a voltage is applied to the electrodes. In equation (8), the expressions of $I^{(n)}$ can be written as functions of the coefficients $\alpha_{abc}$ and $\beta_{abcd}$ as:

$$\begin{aligned}
I^{(1)} &= \sum_a \text{Im}\left[(\beta_{axyy} - \beta_{ayxy})F_{11a} + (\beta_{ayxx} - \beta_{axyx})F_{12a}\right], \\
I^{(2)} &= \sum_a \text{Re}\left[(\beta_{axxy} + |\sigma|^2 \beta_{ayyy})F_{11a} - (\beta_{axxx} + |\sigma|^2 \beta_{ayyx})F_{12a}\right], \\
I^{(3)} &= \sum_a \{\text{Im}[(\alpha_{axy} - \alpha_{ayx})F_{00a}] + \text{Re}[(\beta_{ayxx} - \beta_{axyx})F_{21a} - (\beta_{axyy} - \beta_{ayxy})F_{22a}]\}, \\
I^{(4)} &= \sum_a \{\text{Re}[(\alpha_{axy} + \alpha_{ayx})F_{00a}] + \text{Im}[(\beta_{axxx} + |\sigma|^2 \beta_{ayyx})F_{21a} + (\beta_{axxy} + |\sigma|^2 \beta_{ayyy})F_{22a}]\}, \\
I^{(5)} &= \sum_a \text{Re}\left[(\beta_{axyy} + \beta_{ayxy})F_{11a} - (\beta_{axyx} + \beta_{ayxx})F_{12a}\right], \\
I^{(6)} &= \sum_a \{\text{Re}[(\alpha_{axx} + |\sigma|^2 \alpha_{ayy})F_{00a}] + \text{Im}[(\beta_{axyx} + \beta_{ayxx})F_{21a} + (\beta_{axyy} + \beta_{ayxy})F_{22a}]\}.
\end{aligned} \tag{9}$$

with

$$F_{jla} = \int dr f_j(\rho, z) g_l(\theta) e_a(\boldsymbol{r}) \tag{10}$$

A successful OAM device requires that there exist nonzero components of $F_{1la}$ for $l = 1,2$ and $a = x, y$. Besides $f_j$ and $g_l$, $F_{1la}$ is also affected by the auxiliary weighting field $\boldsymbol{e}(\boldsymbol{r})$, which is related to the geometry of the electrodes, so $F_{1la}$ needs to be optimized for better collection performance and favorable OPGE responses. Not all electrodes can provide nonzero values of $F_{jla}$. A counterexample is the conventional two-terminal rectangular electrodes geometry, with $\boldsymbol{e}(\boldsymbol{r})$ being a constant vector, the value of $F_{1la}$ is zero because the angle integrations of $g_1(\theta) = \cos\theta$ and $g_2(\theta) = \sin\theta$ are zero; thus no OPGE current can be observed, which magnifies the critical role of the geometry of the electrodes, as discussed in the next section. In principle, all $F_{jla}$ values are functions of the OAM order $m$; however, for a suitable electrode geometry and spot size of the LG beam, all $F_{jla}$ values are very weakly dependent on $m$[40]. This property is very important for accurately extracting different OAM orders in experiments, and it makes the measured current signal $I(m, \sigma)$ shows a simple linear dependence on $m$. For example, there are three terms, $m\sigma_i I^{(1)}$, $mI^{(2)}$, and $m\sigma_r I^{(5)}$, linearly dependent on $m$. As long as any of the terms $I^{(1)}$, $I^{(2)}$, or $I^{(5)}$ are nonzero, the photocurrent signal $I(m,\sigma)$ should show a linear dependence on $m$, which can be used to detect the OAM order. In all experimental works demonstrated so far, the OPGE signal is extracted from a CPGE measurement. Under this scheme, the current difference between opposite circularly polarized ($\sigma = \pm i$) LG beams is measured, and the result is given by:

$$I^{\text{CPGE}}(m) = \frac{1}{2}[I^{\text{dc}}(m,i) - I^{\text{dc}}(m,-i)] = mI^{(1)} + I^{(3)} \qquad (11)$$

Furthermore, the usual OPGE current can be separated by the term $mI^{(1)}$:

$$I^{\text{OPGE}} = \frac{1}{2}[I^{\text{CPGE}}(m) - I^{\text{CPGE}}(-m)] = mI^{(1)} \qquad (12)$$

which is proportional to $m$. Therefore, it is $mI^{(1)}$ that dominates the CPGE response and it is used for the OAM detection scheme based on CPGE extraction, and all other currents $I^{(n)}$ for $n = 2,3,\cdots,6$ behave as background currents. As the OPGE current is extracted from the CPGE current, $I^{(3)}$ is the main background current, which should be minimized to reduce its interference to the measurement of $mI^{(1)}$. Now the possibility of detection of the OAM order through CPGE extraction is converted to a problem to determine the situation in which the value of $I^{(1)}$ is nonzero while minimizing the background from other terms at the same time. In principle, the currents from the terms $mI^{(2)}$ and $m\sigma_r I^{(5)}$ can also be treated as the signal for OAM detection, but they have to be extracted by changing the degree of linear polarization, which remains an interesting opportunity to explore experimentally.

To date, the OPGE response and related direct OAM detection device have been experimentally demonstrated on three different materials: WTe$_2$, TaIrTe$_4$, and multilayer graphene (MLG)[32,40,42]. Among them, WTe$_2$ and TaIrTe$_4$ both belong to the C$_{2v}$ point group and share the same symmetry analysis, while MLG belongs to the D$_{6h}$ point group. In either case, the electrical dipole response, which is contributed by the third-rank tensor $\alpha_{abc}$, vanishes under normal incidence conditions due to the lack of planar components $\alpha_{abc}$ for C$_{2v}$ and the existence of inversion symmetry of D$_{6h}$. For C$_{2v}$, there are 21 independent nonzero fourth-rank tensors $\beta_{abcd}$, including eight planar components ( $\beta_{xxxx}, \beta_{xxyy}, \beta_{xyxy}, \beta_{xyyx}$ and $\beta_{yyyy}, \beta_{yyxx}, \beta_{yxyx}, \beta_{yxxy}$ ) and 13 nonplanar components ( $\beta_{zzzz}$ , $\beta_{zzxx}, \beta_{zxzx}, \beta_{zxxz}$ , $\beta_{xxzz}, \beta_{xzxz}, \beta_{xzzx}$ , $\beta_{zzyy}, \beta_{zyzy}, \beta_{zyyz}$ , and $\beta_{yyzz}, \beta_{yzyz}, \beta_{yzzy}$). With such symmetry, devices based on these WTe$_2$ and TaIrTe$_4$ materials are OAM sensitive for both the U-shaped electrodes and starfish-shaped electrodes[32,40,42]. For D$_{6h}$, the 21 nonzero tensor components are the same as those of C$_{2v}$, but only 10 of them are independent (see Table 3). With such relationships between different tensor components, devices based on MLG and starfish-shaped electrodes become OAM insensitive. Therefore, a full symmetry analysis is necessary for the realization of direct OAM detection. Based on Eqn (9), the detected OPGE current signals are determined jointly by the effects of three factors: (i) the geometry of the electrodes appearing in $F_{jla}$, (ii) the response coefficients $\alpha_{abc}$ and $\beta_{abcd}$, which are determined by the electronic properties of the functional material, and (iii) the polarization of the incident LG beams $(m,\sigma)$. Here, the elliptical polarization $\sigma$ in factor (iii) is treated as an adjustable parameter for detecting $m$, and factors (i) and (ii) will be discussed in sequence as follows.

## Symmetry analysis for the geometry of electrodes

| $\hat{\mathbf{e}}(\mathbf{r})$ | type | collection region | $\begin{pmatrix} G_{0x} \\ G_{0y} \end{pmatrix}$ | $\begin{pmatrix} G_{1x} \\ G_{1y} \end{pmatrix}$ | $\begin{pmatrix} G_{2x} \\ G_{2y} \end{pmatrix}$ |
|---|---|---|---|---|---|
| radial $\begin{pmatrix} \cos\theta \\ \sin\theta \end{pmatrix}$ | $U_0$ | $S^\rho(\theta_0, \phi_0)$ | $2\sin\frac{\phi_0}{2}\begin{pmatrix} \cos\theta_0 \\ \sin\theta_0 \end{pmatrix}$ | $\begin{pmatrix} c_1 \\ c_2 \end{pmatrix}$ | $\begin{pmatrix} c_2 \\ c_3 \end{pmatrix}$ |
| | $U_1$ | $S^\rho(\theta_0, \pi)$ | $2\begin{pmatrix} \cos\theta_0 \\ \sin\theta_0 \end{pmatrix}$ | $\begin{pmatrix} \frac{\pi}{2} \\ 0 \end{pmatrix}$ | $\begin{pmatrix} 0 \\ \frac{\pi}{2} \end{pmatrix}$ |
| | $U_2$ | $S^\rho(\theta_0, 2\pi)$ | $\begin{pmatrix} 0 \\ 0 \end{pmatrix}$ | $\begin{pmatrix} \pi \\ 0 \end{pmatrix}$ | $\begin{pmatrix} 0 \\ \pi \end{pmatrix}$ |
| | $U_3$ | $\sum_{l=0}^{1} S^\rho(l\pi + \theta_0, \phi_0)$ | $\begin{pmatrix} 0 \\ 0 \end{pmatrix}$ | $\begin{pmatrix} 2c_1 \\ 2c_2 \end{pmatrix}$ | $\begin{pmatrix} 2c_2 \\ 2c_3 \end{pmatrix}$ |
| | $U_4$ | $\sum_{l=0}^{3} S^\rho\left(\frac{l\pi}{2} + \theta_0, \phi_0\right)$ | $\begin{pmatrix} 0 \\ 0 \end{pmatrix}$ | $\begin{pmatrix} 2\phi_0 \\ 0 \end{pmatrix}$ | $\begin{pmatrix} 0 \\ 2\phi_0 \end{pmatrix}$ |
| azimuthal $\begin{pmatrix} -\sin\theta \\ \cos\theta \end{pmatrix}$ | $S_0$ | $S^\theta(\theta_0, \phi_0)$ | $2\sin\frac{\phi_0}{2}\begin{pmatrix} -\sin\theta_0 \\ \cos\theta_0 \end{pmatrix}$ | $\begin{pmatrix} -c_2 \\ c_1 \end{pmatrix}$ | $\begin{pmatrix} -c_3 \\ c_2 \end{pmatrix}$ |
| | $S_1$ | $\sum_{l=0}^{1} S^\theta(l\pi + \theta_0, \phi_0)$ | $\begin{pmatrix} 0 \\ 0 \end{pmatrix}$ | $\begin{pmatrix} -2c_2 \\ 2c_1 \end{pmatrix}$ | $\begin{pmatrix} -2c_3 \\ 2c_2 \end{pmatrix}$ |
| | $S_2$ | $\sum_{l=0}^{3} S^\theta\left(\theta_0 + \frac{l\pi}{2}, \phi_0\right)$ | $\begin{pmatrix} 0 \\ 0 \end{pmatrix}$ | $\begin{pmatrix} 0 \\ 2\phi_0 \end{pmatrix}$ | $\begin{pmatrix} -2\phi_0 \\ 0 \end{pmatrix}$ |
| $c_1 = \frac{1}{2}(\phi_0 + \cos 2\theta_0 \sin\phi_0),\quad c_2 = \frac{1}{2}\sin 2\theta_0 \sin\phi_0,\quad c_3 = \frac{1}{2}(\phi_0 - \cos 2\theta_0 \sin\phi_0)$ ||||||

Table 2 The factors $G_{la}$ with $l = 0,1,2$ and $a = x, y$ for different types of electrodes listed in Figure 2.

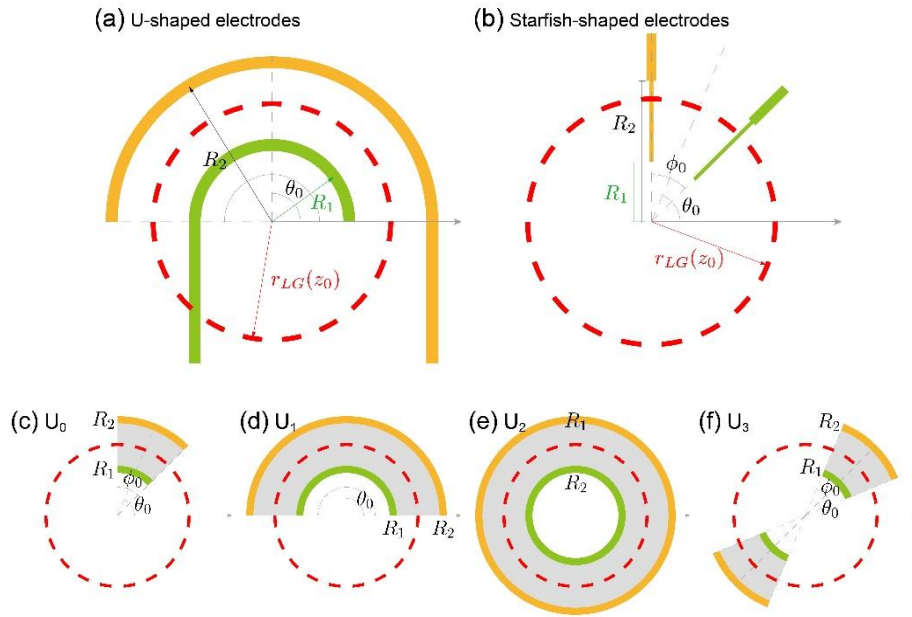

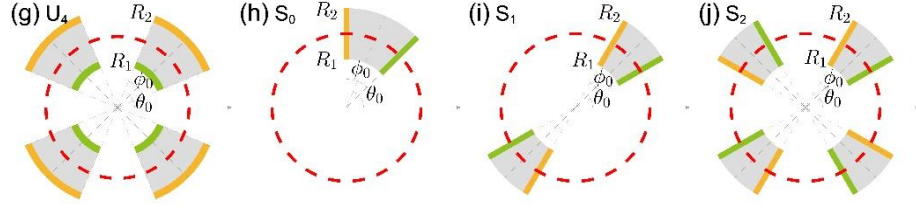

**Fig. 2: Schematic of different electrode designs.** Illustration of the experimentally used U-shaped electrodes (a) and starfish-shaped electrodes (b). The LG beams with different OAM orders are focused to the same ring radius $r_{LG}(z_0)$, represented by the red dashed line. (c-j) Illustration of different types of electrodes. For each type of electrodes, the yellow bars denote one end, and the green bars denote the other end of a pair of electrodes that are connected together, and the shadow region indicates the region where the current is collected from. (c-j) Illustration of different types of electrodes used to collect radial currents (c-g) and azimuthal currents (h-j). (c) The $U_0$ electrodes that collect current from the region formed by $S^\rho(\theta_0, \phi_0)$, which is an intersection between a sector confined by two angles $\theta_0 \pm \phi_0/2$ and an annulus confined by two radii $R_1, R_2$. (d) The $U_1$ electrodes that collect current from the region formed by $S^\rho(\theta_0, \pi)$. (e) The $U_2$ electrodes that collect current from the region formed by $S^\rho(\theta_0, 2\pi)$. (f) The $U_3$ electrodes that collect current from two regions formed by $S^\rho(\theta_0, \phi_0)$ and $S^\rho(\pi + \theta_0, \phi_0)$. (g) The $U_4$ electrodes that collect current from four regions formed by $S^\rho(\theta_0, \phi_0)$, $S^\rho\left(\frac{\pi}{2} + \theta_0, \phi_0\right)$, $S^\rho(\pi + \theta_0, \phi_0)$, and $S^\rho\left(\frac{3\pi}{2} + \theta_0, \phi_0\right)$. (h) The $S_0$ electrodes that collect current from the region formed by $S^\theta(\theta_0, \phi_0)$. (i) The $S_1$ electrodes that collect current from the region formed by $S^\theta(\theta_0, \phi_0)$ and $S^\theta(\pi + \theta_0, \phi_0)$. (j) The $S_2$ electrodes that collect current from the region formed by four regions: $S^\theta(\theta_0, \phi_0)$, $S^\theta\left(\frac{\pi}{2} + \theta_0, \phi_0\right)$, $S^\theta(\pi + \theta_0, \phi_0)$, and $S^\theta\left(\frac{3\pi}{2} + \theta_0, \phi_0\right)$.

Importantly, the peculiarly designed electrodes play a key role in the collection of OAM-sensitive currents based on the idea that the generated photocurrents in the whole laser spot are collected in an unequal weight. For the OPGE detectors that have already been demonstrated experimentally, the electrodes are usually designed into two types: (1) U-shaped electrodes, as shown in Fig. 2 (a), for collecting the photocurrent along the radial direction in the shadow region, and (2) Starfish-shaped electrodes, as shown in Fig. 2 (b), for collecting the photocurrent along the azimuthal direction in the shadow region. The shape and geometry of the electrodes affect the quantity $F_{jla}$ according to Eqn. (10), which depends on the auxiliary weight field $e(r)$. For both types of electrodes, $e(r)$ is distributed mainly in the shadow region from where the photocurrent is collected. This region can generally be represented as an intersection $S(\theta_0, \phi_0)$ between a sector confined by two angles $\theta_0 \pm \frac{\phi_0}{2}$ and an annulus confined by two radii $R_1, R_2$. From Fig. 2 (a, b), the U-shaped electrodes collect the photocurrent from the region formed by $S(\theta_0, \pi) = S^\rho(\theta_0, \pi)$ with $e(r) = e^\rho(r) \equiv \cos\theta\, \hat{x} + \sin\theta\, \hat{y}$, and the starfish-shaped electrodes collect the photocurrent from the region formed by $S(\theta_0, \phi_0) = S^\theta(\theta_0, \phi_0)$ and $e(r) = e^\theta(r) \equiv -\sin\theta\, \hat{x} + \cos\theta\, \hat{y}$; here, we use the superscript $\rho, \theta$ to indicate the two types of electrodes. The factor $F_{jla}$ in Eqn. (10) can be written as $F_{jla} = P_j G_{la}^{\rho/\theta}$, where

$$P_j = \int dz \int_{R_1}^{R_2} \rho\, f_j(\rho, z) d\rho, \quad G_{la}^{\rho/\theta} = \int_{\theta_0 - \frac{\phi_0}{2}}^{\theta_0 + \frac{\phi_0}{2}} d\theta\, g_l(\theta) \hat{e}_a^{\rho/\theta}(r) \tag{13}$$

Here, $\hat{e}_a^\rho(r)$ (or $\hat{e}_a^\theta(r)$) for $a = x, y$ is the $a$th component of the direction of the auxiliary weight field $e^\rho(r)$ (or $e^\theta(r)$). If the values of $F_{1la}$ are all zero, the value of $I^{(1)}$ becomes zero, and the device becomes OAM insensitive regardless of the specific symmetry of the functional material used. In fact, this is indeed the case for conventional two-terminal rectangular electrodes. Thus, it is important to analyze $F_{jla}$ for electrodes with different shapes and symmetries. In general, $P_j$ for $j = 0,1,2$ depends on the radial profile of the LG beam and is not zero, whereas $G_{la}$ for $l = 0,1,2$ and $a = x, y$ depends on the geometry parameters $\theta_0$ and $\phi_0$, which can be strongly affected by the electrodes. The U-shaped and starfish-shaped electrodes used in the experiments have very low symmetry. In order to understand whether electrode symmetry can affect the OPGE current, more types of electrodes with higher symmetries are designed, as shown in Fig. 2 (c-j), for collecting the radial and azimuthal currents. In Table 2, we list the values of $G_{la}$ for different types of electrodes, and below, we present a detailed discussion of how electrode symmetry affects the dependence of the photocurrent on the material's response coefficients:

(1) For the electrodes $U_0$ in Fig. 2 (c), the radial photocurrent is collected in the region $S^\rho(\theta_0, \phi_0)$. When $\theta_0$ and $\phi_0$ take general values, the electrodes do not possess any symmetry, and $G_{la}^\rho$ are all nonzero. In this case, whether the device is OAM sensitive is fully determined by the properties of $\alpha_{abc}$ and $\beta_{abcd}$, and the OAM-sensitive current $I^{(1)}$ depends on four material parameters: $\beta_{axyy} - \beta_{ayxy}$ and $\beta_{ayxx} - \beta_{axyx}$ with $a = x,y$. When the electrodes are placed at certain angles, for example, $\theta_0 = 0$, $G_{0y}^\rho = 0$ and $G_{1y}^\rho = G_{2x}^\rho = 0$, some of the tensor components appearing in Eqn. (9), $\alpha_{abc}$ and $\beta_{abcd}$, will no longer affect the values of $I^{(n)}$. Taking the OAM-sensitive current $I^{(1)}$ as an example, it depends on the two material parameters $\beta_{xxyy} - \beta_{xyxy}$ and $\beta_{yyxx} - \beta_{yxyx}$ and not on $\beta_{yxyy} - \beta_{yyxy}$ and $\beta_{xyxx} - \beta_{xxyx}$. Therefore, the orientation of the electrodes $U_0$ can effectively change the dependence of the OPGE current on the response coefficients.

For the experimentally used U-shaped electrodes in Fig. 2 (a), which are equivalent to the electrodes $U_1$ in Fig. 2 (d), they collect the radial photocurrent in the region $S^\rho(\theta_0, \pi)$. This type of electrodes has mirror symmetry along the direction $\theta_0$, and the coefficients $G_{la}^\rho$ satisfy $G_{1x}^\rho = G_{2y}^\rho = \pi/2$ and $G_{1y}^\rho = G_{2x}^\rho = 0$; in this case, the current $I^{(n)}$ depends on many fewer material parameters because of the mirror symmetry: $I^{(1)}$ only depends on one parameter $\text{Im}[\beta_{xxyy} - \beta_{xyxy} + \beta_{yyxx} - \beta_{yxyx}]$, and $I^{(3)}$ depends on $\alpha_{axy} - \alpha_{ayx}$ and $\beta_{xyxx} - \beta_{xxyx} - \beta_{yxyy} + \beta_{yyxy}$. The advantages of such symmetry analysis can be seen as follows: in order to suppress the CPGE background current $I^{(3)}$, the inversion symmetric materials in class $E_2$ (in Table 3) are also candidates for electrodes $U_1$ for OAM detection, while for electrodes $U_0$, such materials do not work, as will be extensively discussed later in the section "Crystal symmetry analysis for photocurrents $I^{(n)}$ with $n = 1,2,\cdots,6$". Therefore, analyzing the symmetry of electrodes is indispensable for selecting functional materials.

(2) For the starfish-shaped electrodes in Fig. 2 (b), which are equivalent to the $S_0$ type of electrodes in Fig. 2 (h), they collect the azimuthal photocurrent in the region $S^\theta(\theta_0, \phi_0)$. For this type of electrodes, the angle $\phi_0$ has to be small enough to confine the auxiliary weight field inside the shadow region; thus, such types of electrodes impose no additional symmetry and the symmetry cannot be increased simply by changing $\phi_0$. Similar to the $U_0$ electrodes, the OAM-sensitive current $I^{(1)}$ depends on four material parameters: $\beta_{axyy} - \beta_{ayxy}$ and $\beta_{ayxx} - \beta_{axyx}$ with $a=x,y$.

(3) The electrodes can also be designed with other symmetries. The highest symmetry for $U_0$-type of electrodes can be constructed with $\phi_0 = 2\pi$ [45] when the electrodes become annuluses, as illustrated by the $U_2$ electrodes shown in Fig. 2 (e). It has circular symmetry, and these coefficients become $G_{1x}^\rho = G_{2y}^\rho = \pi$ and $G_{1y}^\rho = G_{2x}^\rho = G_{0x}^\rho = G_{0y}^\rho = 0$. In this case, the currents become:

$$\begin{aligned}
I_\rho^{(1)} &= \text{Im}[\beta_{xxyy} - \beta_{xyxy} + \beta_{yyxx} - \beta_{yxyx}]P_1\frac{\phi_0}{2}, \\
I_\rho^{(2)} &= \text{Re}[\beta_{xxxy} + |\sigma|^2\beta_{xyyy} - \beta_{yxxx} - |\sigma|^2\beta_{yyyx}]P_1\frac{\phi_0}{2}, \\
I_\rho^{(3)} &= \text{Re}[(\beta_{xyxx} - \beta_{xxyx} - \beta_{yxyy} + \beta_{yyxy})]P_2\frac{\phi_0}{2}, \\
I_\rho^{(4)} &= \text{Im}[(\beta_{xxxx} + |\sigma|^2\beta_{xyyx} + \beta_{yxxy} + |\sigma|^2\beta_{yyyy})]P_2\frac{\phi_0}{2}, \\
I_\rho^{(5)} &= \text{Re}[\beta_{xxyy} + \beta_{xyxy} - \beta_{yxyx} - \beta_{yyxx}]P_1\frac{\phi_0}{2}, \\
I_\rho^{(6)} &= \text{Im}[(\beta_{xxyx} + \beta_{xyxx} + \beta_{yxyy} + \beta_{yyxy})]P_2\frac{\phi_0}{2}.
\end{aligned} \quad (14)$$

It can be seen that each $I^{(n)}$ depends on only one independent material parameter formed by the planar response tensor components $\beta$ and is independent of $\alpha$. For example, the OAM-sensitive current $I_\theta^{(1)}$ is only determined by $\text{Im}[\beta_{yxyy} - \beta_{yyxy} - \beta_{xyxx} + \beta_{xxyx}]$. In contrast, for electrodes without additional symmetry, $I_\theta^{(1)}$ is determined by four independent material parameters: $\text{Im}[\beta_{xxyy} - \beta_{xyxy}]$, $\text{Im}[\beta_{xyxx} - \beta_{xxyx}]$, $\text{Im}[\beta_{yxyy} - \beta_{yyxy}]$, and $\text{Im}[\beta_{yyxx} - \beta_{yxyx}]$. In addition, we note that similar results in Eq. (14) can be achieved in electrodes with lower symmetry, for example, the $U_4$ electrodes that collect the photocurrent in the region formed by $\sum_{l=0}^{3} S^\rho \left( \frac{l\pi}{2} + \theta_0, \phi_0 \right)$ in Fig. 2 (g), which has $C_4$ rotational symmetry.

(4) Based on the starfish-shaped electrodes, we can also design electrodes with $C_4$ rotational symmetry, as indicated by the $S_2$ electrodes, which collect the photocurrent from the region formed by $\sum_{l=0}^{3} S^\theta \left( \frac{l\pi}{2} + \theta_0, \phi_0 \right)$ in Fig. 2 (j), leading to $G_{0x}^\theta = G_{0y}^\theta = G_{1x}^\theta = G_{2y}^\theta = 0$ and $G_{1y}^\theta = -G_{2x}^\theta = 2\phi_0$. The collected azimuthal currents are as follows:

$$\begin{aligned}
I_\theta^{(1)} &= \text{Im}[\beta_{yxyy} - \beta_{yyxy} - \beta_{xyxx} + \beta_{xxyx}]F_1 2\phi_0, \\
I_\theta^{(2)} &= \text{Re}[\beta_{yxxy} + |\sigma|^2 \beta_{yyyy} + \beta_{xxxx} + |\sigma|^2 \beta_{xyyx}]F_1 2\phi_0, \\
I_\theta^{(3)} &= \text{Re}[(\beta_{yyxx} - \beta_{yxyx} + \beta_{xxyy} - \beta_{xyxy})F_2]2\phi_0, \\
I_\theta^{(4)} &= \text{Im}[(\beta_{yxxx} + |\sigma|^2 \beta_{yyyx} - \beta_{xxxy} - |\sigma|^2 \beta_{xyyy})F_2]2\phi_0, \\
I_\theta^{(5)} &= \text{Re}[\beta_{yxyy} + \beta_{yyxy} + \beta_{xxyx} + \beta_{xyxx}]F_1 2\phi_0, \\
I_\theta^{(6)} &= \text{Im}[(\beta_{yxyx} + \beta_{yyxx} - \beta_{xxyy} - \beta_{xyxy})F_2]2\phi_0.
\end{aligned} \quad (15)$$

Similar to Eq. (14), here, each $I_\theta^{(n)}$ depends only on one independent material parameter formed by the planar response tensor components $\beta$ and is independent of $\alpha$, and the OAM-sensitive current $I_\theta^{(1)}$ is determined by $\text{Im}[\beta_{yxyy} - \beta_{yyxy} - \beta_{xyxx} + \beta_{xxyx}]$. For many crystal symmetries, $I_\theta^{(1)}$ is zero, as shown in Table 3, and this type of electrodes is OAM insensitive under the current OAM detection strategy. However, as discussed later in the perspective section, an alternative strategy based on $I_\theta^{(2)}$ can be used for OAM detection in this case.

(5) When the electrodes have inversion symmetry, such as the $U_2$, $U_4$, and $S_2$ electrodes, as well as the $U_3$ electrodes, which collect photocurrent from the region formed by $\sum_{l=0}^{1} S^\rho (l\pi + \theta_0, \phi_0)$ and the $S_1$ electrodes, which collect photocurrent from the region formed by $\sum_{l=0}^{1} S^\theta (l\pi + \theta_0, \phi_0)$, the inversion symmetry leads to $G_{0x} = G_{0y} = 0$, and the photocurrent signals collected from the conventional second-order photogalvanic effects vanish, regardless of whether the functional material itself possesses inversion symmetry. Therefore, the current from the dipole interaction can be effectively eliminated by a careful electrode design, which provides an effective way to suppress the background signal in OAM direct measurement.

According to the above discussion, the additional symmetry of the electrode geometry can change the dependence of the photocurrent on the response tensor components of $\alpha_{abc}$ and $\beta_{abcd}$ by setting different values of coefficients $F_{lja}$: (1) For electrodes with inversion symmetry, the symmetry of the electrodes imposes $F_{0la} = 0$, which helps eliminate the conventional second-order photogalvanic effects. This would save a large category of materials that breaks inversion symmetry for OAM detection through OPGE because it provides a practical method to reduce the background signal from conventional second-order photogalvanic effects. (2) When the electrodes possess a symmetry higher than $C_4$ rotational symmetry, only two of the four quantities $F_{j1x}$, $F_{j1y}$, $F_{j2x}$, and $F_{j2y}$ are nonzero with the same value; thus, all the signals are from the electric-quadrupole and magnetic-dipole contributions, and each term of $I^{(n)}$ depends only on one independent material parameter, which can be set to zero more easily by choosing a suitable crystal symmetry of the functional material. This approach will be particularly useful in suppressing the background current and increasing the variety of OAM-sensitive materials for OAM detection.

**Crystal symmetry analysis for photocurrents $I^{(n)}$ with $n = 1, 2, \cdots, 6$**

According to the previous discussion, the symmetry from the electrodes has very limited effects on determining whether the values of $I^{(n)}$ with $n = 1,2,\cdots,6$ are zero, but it can effectively suppress the background current from the conventional second-order response tensor $\alpha_{abc}$ by designing electrodes with inversion symmetry and effectively simplify the relationship between each current $I^{(n)}$ and the response tensor components of $\beta_{abcd}$, as shown in Eqns. (9, 14, 15). For normally incident OAM light, only the planar components of $\beta_{abcd}$ are involved. As a fourth-rank tensor, all the components of $\beta_{abcd}$ cannot be zero simultaneously, and its nonzero components determine the nonzero terms of $I^{(n)}$. When the symmetry of the crystal structure is considered, the dependence of the photocurrents $I^{(n)}$ on the response coefficients $\alpha_{abc}$ and $\beta_{abcd}$ can be greatly simplified, and the minimal symmetry requirement to provide OAM sensitivity can be obtained by analyzing Eqns. (9, 14, 15) for different crystal systems. With such symmetry analysis, one can obtain knowledge of the class of materials that are OAM sensitive, and the results are listed in Table 3.

In the table, the nonvanishing planar components of $\alpha_{abc}$ and $\beta_{abcd}$ are listed, as well as the corresponding nonvanishing current component $I^{(l)}_{\rho/\theta}$ based on Eqns. (9, 14, 15), for different crystal systems. Here, the LG beam is incident perpendicularly to the *a-b* plane in the coordinates defined by the crystal axes; a similar analysis can be performed for the light incidence along other directions. All the crystal systems are categorized into four classes ($E_1$, $E_2$, $E_3$, and $E_4$) according to the nonzero planar components of $\beta_{abcd}$. The $E_1$ class includes isotropic, hexagonal ($D_6$, $C_{6v}$, $D_{6h}$), trigonal ($D_{3d}$, $D_3$, $C_{3v}$), cubic (O, $T_d$, $O_h$) and tetragonal ($D_4$, $C_{4v}$, $D_{4h}$, $D_{2d}$) crystals. In the $E_1$ class, only $I^{(1)}_\rho$, $I^{(4)}_\rho$, $I^{(2)}_\theta$, and $I^{(3)}_\theta$ can be nonzero for the electrodes $U_2$, $U_4$, and $S_2$ with the highest symmetry; the radial signal is $I(m,\sigma) = m\sigma_i I^{(1)}_\rho + I^{(4)}_\rho$, and the azimuthal signal is $I(m,\sigma) = m I^{(2)}_\theta + \sigma_i I^{(3)}_\theta$. For $U_1$ electrodes, the radial signal is $I(m,\sigma) = m\sigma_i I^{(1)}_\rho + I^{(4)}_\rho + \sigma_r I^{(6)}_\rho$, but $I^{(6)}_\rho$ arises from the conventional photogalvanic effects only. For the electrodes $U_0$, $U_3$, $S_0$, and $S_1$, the symmetry is much lower, and more terms of $I^{(n)}$ become nonzero, but $I^{(1)}_\theta$ is still zero. Therefore, for the existing detection strategy based on CPGE current extraction, only the radial current ($I^{(1)}$) can be collected with the background CPGE current $I^{(3)} = 0$, but no OPGE current can be obtained from the azimuthal current. However, because the term $m I^{(2)}_\theta$ is linearly related to the OAM order *m*, in principle, OAM detection is also possible if the term $mI_\theta^{(2)}$ can be extracted from the azimuthal current when adopting appropriate method, as discussed in the last section. The $E_2$ class includes cubic (T, $T_h$), monoclinic ($C_{2h}$, $C_2$, $C_{1h}$), and orthorhombic ($D_2$, $C_{2v}$, $D_{2h}$) crystals. For electrodes $U_0$, $U_3$, $S_0$, and $S_1$ with lower symmetry, all terms of $I^{(n)}$ are nonzero. While for electrodes with higher symmetries, compared to the $E_1$ class materials, the $E_2$ class materials have two additional nonzero photocurrent response terms from $m\sigma_r I_\rho^{(5)}$ and $\sigma_r I_\theta^{(6)}$. For the existing detection strategy based on CPGE current extraction, both the $m\sigma_r I_\rho^{(5)}$ and $\sigma_r I_\theta^{(6)}$ terms do not contribute to the CPGE because the incident OAM beam is circularly polarized with $\sigma_r = 0$. The $E_3$ class includes hexagonal ($C_6$, $C_{6h}$, and $C_{3h}$), trigonal ($S_6$ and $C_3$), and tetragonal ($C_4$, $S_4$, and $C_{4h}$) crystals. For the $E_3$ class, the photocurrent response becomes $I(m, \sigma) = m\sigma_i I^{(1)} + mI^{(2)} + \sigma_i I^{(3)} + I^{(4)}$ for both the radial current and azimuthal current under circularly polarized LG beams. All current components exist for materials in the $E_3$ class, and the response has to be analyzed case by case. While for the $E_4$ class, which includes all the materials that are not categorized into the previous three classes, all the components of $I^{(n)}$ for $n = 1, 2, \cdots, 6$ are nonzero.

Therefore, from a symmetry point of view, after combining the electrodes' symmetry and the crystal symmetry, all materials are OAM sensitive in the radial currents in the conventional OPGE detection scheme; those materials from classes $E_3$ and $E_4$ are OAM sensitive in the azimuthal currents for electrodes with higher symmetries; and materials from the $E_2$ classes can be OAM sensitive in the azimuthal currents for electrodes with much lower symmetries.

| crystal system | nonvanishing planar components | | nonvanishing currents | | | | | class |
|---|---|---|---|---|---|---|---|---|
| | $\alpha_{abc}$ | $\beta_{abcd}$ | $U_0$ $U_3$ | $S_0$ $S_1$ | $U_1$ | $U_2$ $U_4$ | $S_2$ | |
| Isotropic Hexagonal ($D_6$, $C_{6v}$, $D_{6h}$) Trigonal($D_{3d}$, $D_3$) | - | $xxyy = yyxx$ $xyxy = yxyx$ $xyyx = yxxy$ $xxxx = yyyy$ | $I_\rho^{(1)}$ $I_\rho^{(2)}$ $I_\rho^{(4)}$ $I_\rho^{(5)}$ $I_\rho^{(6)}$ | $I_\theta^{(2)}$ $I_\theta^{(3)}$ $I_\theta^{(4)}$ $I_\theta^{(5)}$ $I_\theta^{(6)}$ | $I_\rho^{(1)}$ $I_\rho^{(4)}$ $I_{\rho,*}^{(6)}$ | $I_\rho^{(1)}$ $I_\rho^{(4)}$ | $I_\theta^{(2)}$ $I_\theta^{(3)}$ | $E_1$ |
| Trigonal($C_{3v}$) | $yxx = xxy$ $= xyx$ $= -yyy$ | $xxxx = xxyy +$ $xyxy + xyyx$ | | | | | | |
| Cubic ($O$, $T_d$, $O_h$) Tetragonal ($D_4$, $C_{4v}$, $D_{4h}$, $D_{2d}$) | - | $xxyy = yyxx$ $xyxy = yxyx$ $xyyx = yxxy$ $xxxx = yyyy$ | | | | | | |
| Cubic($T$, $T_h$) | - | $xxyy, yyxx$ $xyxy, yxyx$ $xyyx, yxxy$ $xxxx = yyyy$ | all | all | $I_\rho^{(1)}$ $I_\rho^{(4)}$ $I_\rho^{(5)}$ $I_\rho^{(6)}$ $I_{\rho,*}^{(6)}$ | $I_\rho^{(1)}$ $I_\rho^{(4)}$ $I_\rho^{(5)}$ | $I_\theta^{(2)}$ $I_\theta^{(3)}$ $I_\theta^{(6)}$ | $E_2$ |
| Monoclinic($C_{2h}$) Orthorhombic ($D_2$, $C_{2v}$, $D_{2h}$) | - | $xxyy, yyxx$ $xyxy, yxyx$ $xyyx, yxxy$ $xxxx, yyyy$ | | | | | | |
| Monoclinic($C_2$, $C_{1h}$) | $xxy, xyx,$ $yxx, yyy$ | | | | | | | |
| Hexagonal ($C_6$, $C_{6h}$) Trigonal($S_6$) | - | $xxyy = yyxx$ $xyxy = yxyx$ $xyyx = yxxy$ $xxxx = yyyy$ $yyxy = -xxyx$ $yxyy = -xyxx$ $xyyy = -yxxx$ $xxxy = -yyyx$ | all | all | $I_\rho^{(1)}$ $I_\rho^{(2)}$ $I_\rho^{(3)}$ $I_\rho^{(4)}$ $I_\rho^{(6)}$ | $I_\rho^{(1)}$ $I_\rho^{(2)}$ $I_\rho^{(3)}$ $I_\rho^{(4)}$ | $I_\theta^{(1)}$ $I_\theta^{(2)}$ $I_\theta^{(3)}$ $I_\theta^{(4)}$ | $E_3$ |
| Hexagonal ($C_{3h}$) Trigonal($C_3$) | $xyy = yxy$ $= yyx$ $= -xxx$ $yxx = xyx$ $= xxy$ $= -yyy$ | $xxxx = xxyy +$ $xyxy + xyyx$ $xxxy = yyxy +$ $yxyy + xyyy$ | | | | | | |
| Tetragonal ($C_4$, $S_4$, $C_{4h}$) | - | $xxyy = yyxx$ $xyxy = yxyx$ $xyyx = yxxy$ $xxxx = yyyy$ $yyxy = -xxyx$ $yxyy = -xyxx$ $xyyy = -yxxx$ $xxxy = -yyyx$ | | | | | | |
| Triclinic($C_1$, $S_2$), other | all | all | all | | | | | $E_4$ |

Table 3 Crystal symmetry analysis for the planar components of the response tensors $\alpha_{abc}$ and $\beta_{abcd}$. The crystal system is adopted from Boyd's book[51], and the column "nonvanishing currents" denotes nonzero terms of $I^{(n)}$ for n = 1,2,⋯,6 for several types of electrodes listed in Table 2. For the type $U_1$, the subscript symbol "*" in $I_{\rho,*}^{(6)}$ means that $I_\rho^{(6)}$ is nonzero only when a nonzero second-order response exists.

**Experimental Progress**

Based on the above symmetry analysis, a device made from photosensitive materials with an OPGE response and correct electrode geometry should be able to resolve the OAM order when the circular polarization-dependent photocurrent is extracted from a standard CPGE measurement. The pioneering work was performed by Ji *et al*. using $WTe_2$. The typical device is based on a U-shaped electrode geometry (Fig. 3a), which can effectively collect the radial photocurrent response when the OAM beam is embedded between the inner and outer electrodes. To extract the CPGE response, a quarter wave plate after a linear polarizer is employed and rotated to modulate the polarization of the incident OAM beam. The polarization of the OAM beam undergoes a periodic change of linear (0°)–left circular (45°)–linear (90°)–right circular (135°)–linear (180°) polarization states in an operation cycle of 180°. By measuring the photocurrent with certain degree interval, the CPGE response can be directly extracted via Fourier transform, which corresponds to the 180°-periodicity component of the photocurrent (Fig. 3b). The experimental results show that the extracted CPGE response displays a linear and steplike dependence on the topological charge of the OAM beam, which enables the resolution of the OAM order to range from +4--4 (Fig. 3c). However, the linear polarization–dependent current $J_L$ (90°-periodicity component) and the polarization-independent thermal current $J_0$ do not have such dependence on the OAM order.

Since $WTe_2$ has an OPGE response in both the radial and azimuthal directions, three different electrode geometries are tested, and all the results clearly demonstrate a CPGE response that reverses in sign when the OAM number goes opposite (Fig. 3d-i). However, U-shaped electrode geometries usually demonstrate better performance than other electrode geometries because this geometry is more tolerant of position errors at different OAM orders than the starfish-electrode device is. The OPGE detector based on $WTe_2$ works in the near-infrared region but has a very limited response in the mid-infrared region. Later on, Lai *et al*. replaced $WTe_2$ with $TaIrTe_4$, a type-II Weyl semimetal with exactly the same $C_{2v}$ crystal symmetry as $WTe_2$, realizing OAM detection in the mid-infrared region on the basis of the same device geometry (Fig. 3j-l). This success mainly takes advantage of the topologically enhanced responsivity at the mid-infrared region while maintaining the OPGE response that is allowed by $C_{2v}$ symmetry. In a more recent work by Yang *et al*., mid-infrared OAM detection was also realized with multiple-layer graphene (MLG) with unexpectedly high OAM resolution compared with $TaIrTe_4$ (Fig. 3m-n). Compared with $TaIrTe_4$, the absolute responsivity in the mid-infrared region underscores the advantage of MLG in photoelectric conversion. The improved response is due to two reasons: 1. the crystal symmetry $D_{6h}$ of the MLG only allows OAM sensitive photocurrent collection with U-shaped electrodes, but the CPGE background signal is suppressed with this collection geometry, which suppresses noise from the background; 2. the response coefficients are greatly enhanced due to the low-dimensional effect, which is induced mainly by Drude-type intraband resonant transitions for massless Dirac Fermions. Such dimensionality enhancement only occurs in two dimensions instead of three dimensions, as discussed in detail in the work[42] (Fig. 3o). Because MLG is an epitaxially growable material with high stability in moderate environments and can be easily integrated with the silicon chip technique[52-56], the successful demonstration of an MLG-based OAM detector paves the way for the development of on-chip detectors and large-scale FPA devices (Fig. 1a, 1b).

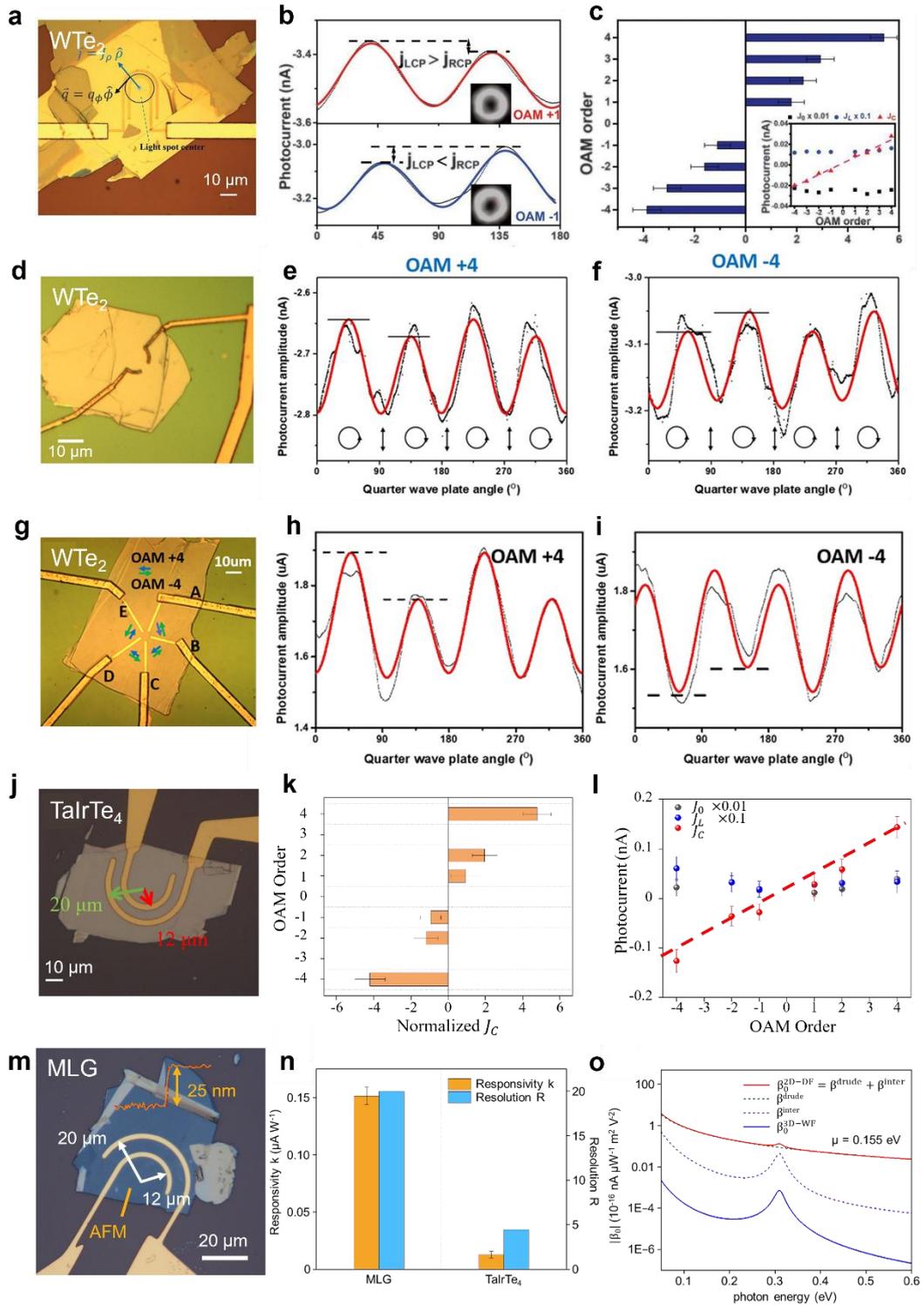

**Fig. 3: Experimental results and progress of OPGE-based OAM photodetection.** (a-c) Experimental results of the U-shaped WTe$_2$ device, including an optical image (a), the measured photocurrent response under the excitation of OAM ±1 beams as a function of the QWP angle (b), and the normalized CPGE response as a function of the OAM order ranging from −4 to 4 (c). (a-i) Reproduced with permission[32], Copyright 2020, American Association for the Advancement of Science. (d-i) Experimental results of Ω-shaped (d-f) and starfish-shaped (g-i) WTe$_2$ devices, including optical images (d, g) and CPGE measurements for

OAM orders +4 (e, h) and -4 (f, i).(j-l) Reproduced with permission[40], Copyright 2022, John Wiley & Sons. Experimental results of the U-shaped TaIrTe$_4$ device, including an optical image (j), normalized CPGE response as a function of the OAM order ranging from −4 to 4 (h), and linear fitting of the CPGE, anisotropic and polarization-independent components of the measured photocurrent as a function of the OAM order (l). (m-o) Reproduced under the terms of the CC-BY Creative Commons Attribution 4.0 International license (https://creativecommons.org/licenses/by/4.0).[42] Copyright 2025, The Authors, published by Springer Nature. Experimental results and analysis of the U-shaped multilayer graphene device, including optical images (m), comparison of the OPGE response and OAM resolution capability with those of the U-shaped TaIrTe$_4$ device (n), and calculations of the OPGE response coefficients for 2D-dirac fermions and 3D-weyl fermions (o).

**Operation speed**

In a typical OPGE response measurement, the circular polarization-dependent component of the photocurrent response must be extracted to distinguish the OAM orders of light. Simultaneous measurement of the left- and right-circular polarization-dependent photocurrents was realized through continuous rotation of a quarter waveplate (QWP) in early works[32,40,42], as illustrated in Fig. 3a. The CPGE component is then extracted via Fourier transform of the QWP angle-dependent photocurrent response. Limited by the speed of mechanical polarization modulation, the operation speed is at the minute level for early demonstration, which cannot fulfill the speed requirements of most applications[32,40,42]. The slow operation speed is a major drawback of OPGE-based OAM detectors compared with parallel OAM detection technology based on SPPs, which can reach operation speeds on the order of tens of microseconds[34]. In a recent work by Yang *et al.*, the operation speed was greatly increased to the mini-second level via an electrical polarization modulation strategy with a photoelastic modulator (PEM) accompanied by a phase locked readout approach with a lock-in amplifier. A detailed schematic diagram of their modulation and measurement scheme is shown in Fig. 3b-c. The PEM used in this work consists of a ZnSe$_2$ resonant bar operating at a resonance frequency of 50.14 kHz, which is capable of mid-infrared circulation polarization modulation. When the OAM beam passes through the PEM, in a single operational cycle of the PEM, the polarization of the OAM beam undergoes a sequence of transitions—linear, left-circular, linear, right-circular, linear, exhibiting a variation pattern of polarization modulation similar to that realized by rotating the QWP over a period of 180° (Fig. 3d-f). In the detection part, with an OPGE-based OAM detector made from multilayer graphene (MLG), the topological charge of light OAM can be clearly distinguished by the quantized plateau of the CPGE response, which is directly extracted by a lock-in amplifier that is phase locked to the modulation signal of the PEM (Fig. 3c). In this work, the readout speed is mainly limited by the 300 ms measurement time constant of the lock-in amplifier (Fig. 3g). The measurement time constant of the lock-in amplifier can be reduced to compensate for the elevated noise level, and their signal-to-noise level barely allows a measurement time constant of 1 ms (Fig. 3h). Providing enough OPGE response of the device, the measurement time of the lock-in can be further decreased, and the operation speed could be improved further under this modulation scheme. The time constant is ultimately limited by the polarization modulation frequency of the PEM and the photocurrent response time of the multilayer graphene device, both of which could be improved further in the future[57-59]. In addition, the electric modulation scheme and lock-in readout scheme are directly applicable to focal plane array devices and on-chip integration (Fig. 3i-l). For the focal plane array device, the modulation crystal with the transducer assembly can be placed before the OAM detector arrays, and the phase-sensitive reading circuit can be integrated with the photocurrent readout circuit and locked to the electric driving source to extract the CPGE component of each detector cell. For on-chip integration, miniaturized electrically driven polarization modulation materials are required for OAM detectors, and this is possible with atomically thin topological semimetals for tunable phase retardation[60].

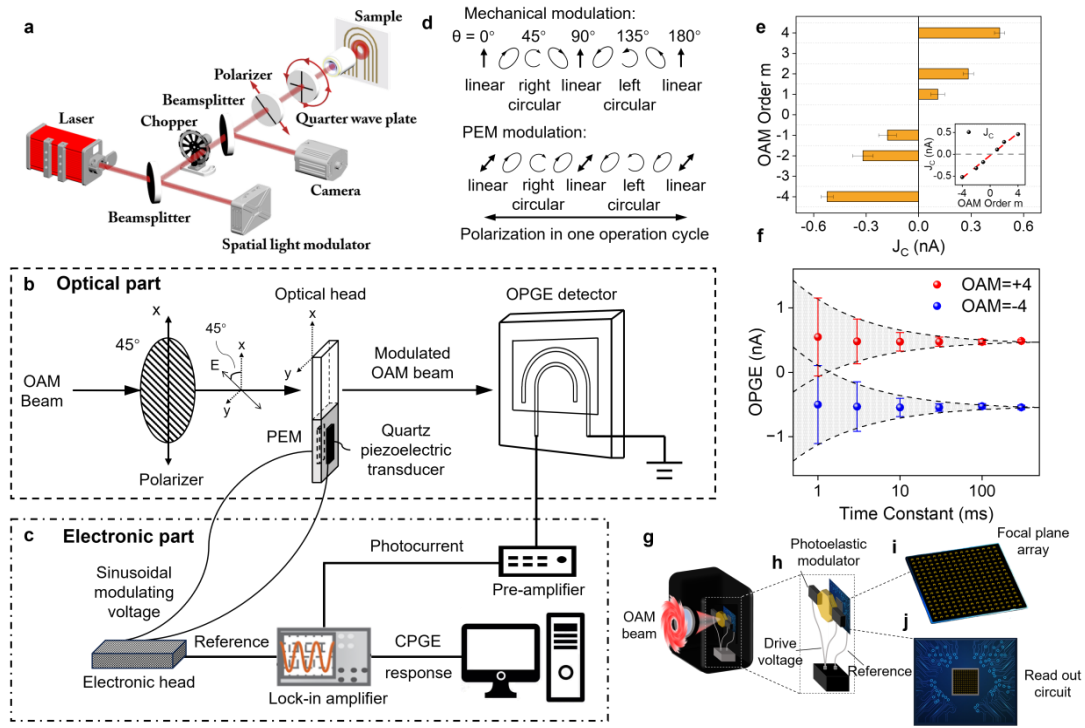

**Fig. 4: Schematic of PEM modulation and high-speed OAM photodetection experiments.** (a) Experimental setup for OPGE measurement based on mechanical modulation. Reproduced with permission[32], Copyright 2020, American Association for the Advancement of Science. (b-c) Experimental setup including an optical part (b), polarization modulation with the polarizer and optical head of the PEM) and an electronic part (c), modulating voltage, photocurrent collection and CPGE extraction) for OPGE measurements based on PEM modulation.[41] (d) Comparison of the polarization variation in one operation cycle between mechanical and PEM modulations.[41] (e) Measured CPGE response $J_C$ based on PEM modulation, together with its linear fit (inset), as a function of the OAM order m.[41] (f) Measured CPGE response together with its uncertainty as a function of the time constant of the lock-in amplifier under the excitation of OAM beams with an OAM order of ±4.[41] (g-j) Schematic of the application of the PEM modulation scheme for light OAM photodetector focal-plane-array devices based on graphene, including the overall device structure of the focal-plane-array device (g), schematic of the OAM detection chip and PEM modulation module driven by the power module (h), the focal-plane array based on MLG photodetectors (i) and the read-out circuit (j).[41] (b-j) Reproduced with permission[41], Copyright 2025, SPIE.

**Challenges and Perspectives**

Experimentally demonstrated direct OAM devices are all based on CPGE measurements, and the extraction of the OPGE current requires the modulation of the polarization states of LG beams, which leads to a complicated detection procedure. In fact, on the basis of our symmetry analysis shown in Table 3, various detection schemes could make the detection procedure simpler. For example, when highly symmetric electrode $S_2$ is used for materials with $E_1$ class symmetry, direct detection based on the photocurrent response $I^{dc}(m,\sigma) = mI^{(2)} + \sigma_i I^{(3)}$ can be performed, and the detected current can be directly quantified with the OAM order $m$ as long as the LG beam is linearly polarized or unpolarized; this direct extraction strategy can greatly improve the detection speed, which is limited by the polarization modulation and related extraction process in the conventional detection scheme. However, experimentally, such a detection scheme may suffer from a photocurrent response that results from trivial effects that are OAM dependent. For example, different beam profiles at different OAM orders can contribute to an OAM-dependent photocurrent response and interfere with detection. Furthermore, detection can also be interfered with by other photocurrent response mechanisms that are OAM sensitive, such as the s-PGE response reported in previous work[61]. In light of these difficulties, more experimental and theoretical studies are expected to explore this new detection scheme.

To date, all discussions are limited to the detection of single scalar OAM beams, and there is a direct correspondence between the OPGE current and topological charge of OAM beams, which enables direct detection of scalar OAM beams. A step further involves the detection of a mixture of different OAM orders, and a protocol for detecting such a mixture was proposed in the first OPGE detector by Ji et al.[32], who suggested the use of a well-designed matrix of electrodes (Fig. 5a) for this purpose. Considering the mixture of n LG modes $\boldsymbol{E}(LG_0^1)\ldots\boldsymbol{E}(LG_0^n)$ with percentages $x_1\ldots x_n$, the mixed light field and OPGE photocurrent intensities are expressed as:

$$\boldsymbol{E}(\rho,\theta)_{mix} = x_1\boldsymbol{E}(LG_0^1) + x_2\boldsymbol{E}(LG_0^2) + \cdots + x_n\boldsymbol{E}(LG_0^n)$$
$$J_{OPGE,mix} \propto \frac{e^{-2\rho^2/w_0^2}}{\rho}\left(\frac{x_1}{|1|!}\rho^2 + \frac{2x_2}{|2|!}\rho^4 + \cdots + \frac{nx_n}{|n|!}\rho^{2n}\right) \quad (16)$$

where $\rho$ is the radial coordinate. As the intensity profiles of different OAM mixture is different, the detection with a series of electrodes designed with different radii can, in principle, distinguish the mixture of modes when accompanied by proper postdata processing.

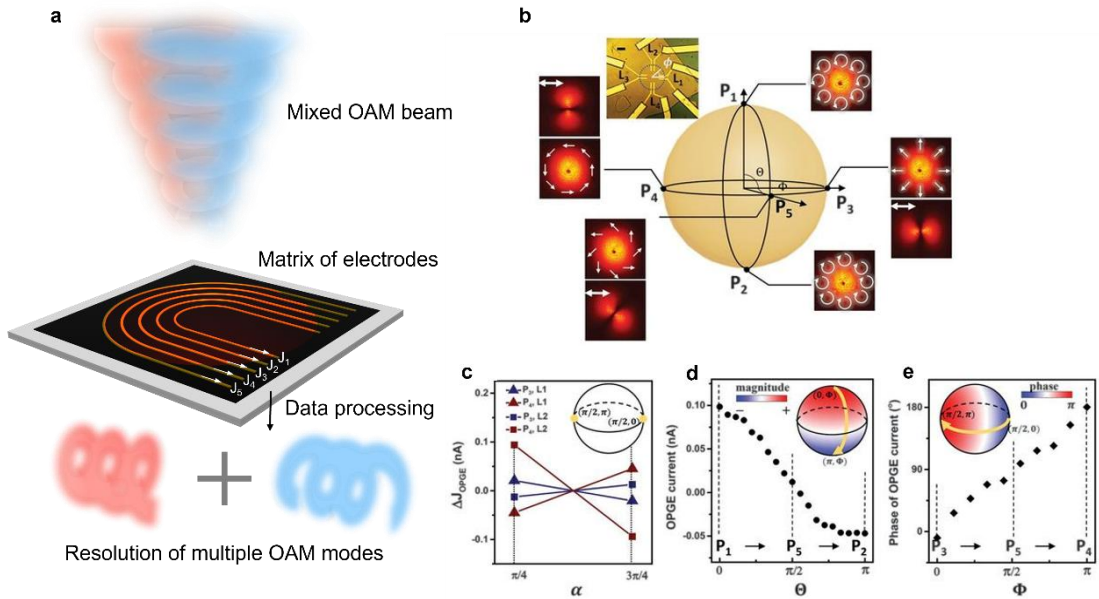

**Fig. 5: Prospects for detecting more complex vortex beams.** (a) Schematic of the device geometry and process for measuring arbitrary mixtures of OAM beams. (b-e) Reproduced with permission[32], Copyright 2020, American Association for the Advancement of Science.Experimental example and prospects for direct detection of a vectorial OAM beam represented on a higher-order Poincaré sphere (HOPS), including the following: (b) Schematic of *m* = 1, σ = −1 HOPS, with states represented by (Θ, Φ) spherical coordinates and five points $P_1$ to $P_5$ together with their polarization distributions with or without a linear polarizer oriented in the horizontal direction and their corresponding intensity profiles. (Inset) Optical image of the octopus-shaped electrodes with four pairs of electrodes ($L_1$, $L_2$, $L_3$, and $L_4$) located at four azimuthal coordinates (ϕ=0, π/2, π, and 3π/2, respectively). (c) Relative photocurrent amplitudes at two QWP angles (π/4 and 3π/4) from two states ($P_3$ and $P_4$) on HOPS measured at two locations ($L_1$ and $L_2$). (d) OPGE current amplitude from a set of states on the line connecting $P_1$, $P_5$, and $P_2$ with the same Φ, and (e) phase of the OPGE current from a set of states on the line connecting $P_3$, $P_5$, and $P_4$ with a fixed Θ.

An even more complicated case involves vectorial OAM beams, which process space-variant states of polarization in addition to the helical phase distribution. This was also discussed in the very first OPGE detector work by Ji *et al.*[32]. Vectorial OAM beams can be represented on a higher-order Poincaré sphere (HOPS)[62,63] (Fig. 5b). In the parameter space (represented by the spherical coordinates Θ and Φ) of the HOPS, the state of the optical field is represented by $|\Psi(\theta,\Phi)\rangle = \cos\left(\frac{\theta}{2}\right)\exp\left(-\frac{i\Phi}{2}\right)|L_{-m}\rangle + \sin\left(\frac{\theta}{2}\right)\exp\left(\frac{i\Phi}{2}\right)|R_m\rangle$, where $|R_m\rangle$ and $|L_{-m}\rangle$ are scaler vortex beams with OAM +m (−m) and SAM σ = −1 (+1), respectively. The OPGE response can be divided into two parts, $J^{(0)}_{\phi,OPGE}$ and $J^{(\phi)}_{\phi,OPGE}$, according to their dependence on the azimuthal angle ϕ, expressed as:

$$J_{\phi,OPGE} = J_{\phi,OPGE}^{(0)} + J_{\phi,OPGE}^{(\phi)}$$
$$\propto \frac{m}{\rho}[(c_0 + c_1 \cos\Theta) + (c_2 \cos(2(m+\sigma)\phi + \Phi) + c_3 \sin(2(m+\sigma)\phi + \Phi))\sin(\Theta)] \quad (17)$$

where $c_0$, $c_1$, $c_2$, and $c_3$ are conductivity coefficients. To capture this azimuthal angle dependence, the electrodes were arranged in an "octopus" shape (Fig. 5b, inset) to enable a set of azimuthal current measurements at various azimuthal coordinates ($L_1$--$L_4$), while the beam was fixed at the center defined by the electrodes. The experimental results show that for both the P3 ($\pi/2$, 0) and P4 ($\pi/2$, $\pi$) states, the difference in the OPGE response at two QWP angles ($\pi/4$ and $3\pi/4$) and $\Delta J_{OPGE}^{(\phi)}$ collected at different azimuthal angles $L_1$ and $L_2$ are of opposite sign, indicating the existence of an $J_{OPGE}^{(\phi)}$ that originates from the vectorial OAM beams (Fig. 5c). Moreover, the experimental results show that the amplitude of the OPGE current directly corresponds to $\Theta$ when $\Phi$ is fixed, and the phase of the OPGE current can be mapped onto the $\Phi$ coordinate on HOPS when $\Theta$ is fixed ($\pi/2$) (Fig. 5d-e), indicating the distinct OPGE response related to the states on the HOPS. On the basis of these results, it can be expected that the OAM order or the coordinates of any arbitrary OAM state on a HOPS can be specifically determined by measuring currents via a small matrix of electrodes with suitable data processing.

With the successful demonstration of these single devices, an important step forward is to integrate them into a chip or FPA devices, which is evitable for practical application[64] such as OAM imaging and multi-dimensional light detection. However, it would be very challenging to design an accurate OAM detector for imaging purpose as well as to develop an efficient method to extract the OAM order. This is because the detection of OAM order in each pixel requires the global information of the light, and the detection accuracy strongly depends on the relative position between OAM light and the collection electrodes. The challenge can be solved by resolving full optical parameters using a deep learning network, which has been successfully demonstrated in a recently developed multifunctional photodetectors[65,66]. By integrating the OAM detector, intensity detector[67,68], and polarization detector[69] as a unit of a FPA devices, the existing non-local high-dimensional photodetector[65] can be further expanded to add OAM sensitivity.

The successful demonstration of graphene for OAM detection with high recognition capability is promising for large-scale FPA device integration. As the properties of graphene are versatilely tunable by an external gate[64], it facilitates intelligent detection through a deep learning network[66]. Following the previous successful demonstration of full-Stokes parameters and wavelength detection through a twisted double bilayer graphene photodetector with a trained convolutional neural network[64], it will be possible to detect the OAM order, intensity, and polarization in one detector using twisted bilayer graphene as a functional material and further scale it up as an FPA working unit in the midinfrared region[64]. Regardless of the scheme adopted, the extension of a single OAM detector to an FPA device is evitable for practical applications[64]. Furthermore, although graphene seems to be a strong candidate from the consideration of many aspects, our symmetry analysis of sensitive materials and related electrode designs will provide more space for the discovery and development of more high-quality materials. In addition, the intrinsic quantum geometries of various quantum materials present many opportunities for OAM detection via

OPGE, together with its integration for full optical parameter detection and large-scale on-chip integration.

## Acknowledgments
This project was supported by the National Natural Science Foundation of China (Grant Nos. 62325401, 62250065 and 12034003), the National Key Research and Development Program of China (Grant Nos. 2021YFA1400100 and 2020YFA0308800), the authors also would like to thank the support from the National Natural Science Foundation of China (Grant Nos. 12034001 and 62227822), and the Open Fund of the State Key Laboratory of Infrared Physics (Grant No. SITP-NLIST-ZD-2023-02).

## Author Declarations
**Conflict of Interest**
The authors have no conflicts to disclose.
**Author Contributions**
**Jinluo Cheng:** Conceptualization (equal); Funding acquisition (equal); Methodology (leading); Project administration (equal); Supervision (equal); Validation (leading); Writing – original draft (equal); Writing – review & editing (equal). **Dehong Yang:** Visualization (equal); Writing – original draft (supporting); Writing –review & editing (supporting). **Weiming Wang:** Methodology (supporting); Visualization (equal); Writing – original draft (supporting). **Chang Xu:** Writing – original draft (supporting); Writing –review & editing (supporting). **Zipu Fan:** Visualization (equal); Writing – original draft (supporting); Writing –review & editing (supporting). **Dong Sun:** Conceptualization (equal); Data curation (equal); Funding acquisition (equal); Project administration (equal); Supervision (equal); Writing – original draft (equal); Writing – review & editing (equal).
**Data availability**
Data sharing is not applicable to this article as no new data were created or analyzed in this study.